\documentstyle[aps,epsf,epsfig,rotate,multicol]{revtex}

\begin{document}

\draft

\title{Identifying Business Sectors from Stock Price Fluctuations}

\author{Parameswaran Gopikrishnan$^{1}$, Bernd Rosenow$^{1,2}$, 
Vasiliki Plerou$^{1,3}$, and~H.~Eugene~Stanley$^{1}$ }

\address{$^{1}$ Center for Polymer Studies and Department of Physics, Boston
University, Boston, Massachusetts 02215, USA \\
$^{2}$ Department of
        Physics, Harvard University, Cambridge, Massachusetts
        02138, USA \\
$^{3}$ Department of
        Physics, Boston College, Chestnut Hill, Massachusetts
        02167, USA \\}

\date{\today}

\maketitle

\begin{abstract}
  	  
  Firms having similar business activities are correlated. We analyze
  two different cross--correlation matrices {\bf \sf C} constructed
  from (i) 30--min price fluctuations of 1000 US stocks for the
  2--year period 1994--95 and (ii) 1--day price fluctuations of 422 US
  stocks for the 35--year period 1962--96. We find that the
  eigenvectors of {\bf \sf C} corresponding to the largest eigenvalues
  allow us to partition the set of all stocks into distinct
  subsets. These subsets are similar to conventionally-identified
  business sectors, and are stable for extended periods of time. Using
  a set of coupled stochastic differential equations, we argue how
  correlations between stocks might arise. Finally, we demonstrate
  that the sectors we identify are useful for the practical goal of
  finding an investment which earns a given return without exposure to
  unnecessary risk.

\end{abstract}
\begin{multicols}{2}

The internal structure of a complex system manifests itself in
correlations among its constituents.  In physical systems, one relates
correlations to basic interactions, but for the stock market
problem~\cite{farmer}, the underlying `interactions' are not
known. Suppose that the change of stock prices can be visualized by
the motion of point particles. Correlated particle motion can be
pictured as ``strings'' connecting pairs of particles. Given only the
records of the particle positions at equal time intervals, how can we
identify the strings without `seeing' them?  One approach is to first
calculate the cross-correlation matrix {\bf \sf C} whose elements
$C_{ij}$ are the correlation-coefficients between the velocities of
two particles $i$ and $j$. The eigenvectors of {\bf
\sf C} convey information about the collective modes of the system.

What is the analog of the cross-correlation matrix {\bf \sf C} for the
stock market problem? We define the cross--correlation matrix {\bf \sf
C} with elements $C_{ij} \equiv [\langle G_i G_j \rangle - \langle G_i
\rangle \langle G_j \rangle]/ \sigma_i \sigma_j$, where $\sigma_i$ is the
standard deviation of price fluctuations $G_i(t)\equiv \ln
S_i(t+\Delta t) - \ln S_i(t)$ (returns), $S_i(t)$ denotes the price of
stock $i=1,\dots,N$, and $\langle\dots\rangle$ denotes a time average
over the period studied. To investigate correlations on different time
scales, we analyze (i) 30-min returns of $N=1000$ largest stocks for
the two-year period 1994-95 and (ii) daily returns of $N=422$ stocks
for the 35-year period 1962-96~\cite{TAQ}.

We first diagonalize {\bf \sf C} and rank-order its eigenvalues
$\lambda_k$ such that $\lambda_{k+1}>\lambda_k$; the corresponding
eigenvectors are denoted {\bf \sf u$^{k}$}. Next, we analyze the
components of those {\it deviating eigenvectors} whose eigenvalues are
larger than the upper bound for uncorrelated time
series~\cite{Laloux99,Plerou99}. A direct examination of these
eigenvectors, however, does not yield a straightforward interpretation
of their economic relevance. To interpret their meaning, we note that
the largest eigenvalue is an order of magnitude larger than the
others, which constrains the remaining $N-1$ eigenvalues since {\rm Tr
{\bf \sf C}}~$=N$. Thus, in order to analyze the contents of the
deviating eigenvectors, we first remove the effect of the largest
eigenvalue~\cite{removem}.

To analyze the information contained in the eigenvectors {\bf \sf
u$^k$}, we partition the 1000 stocks into groups labeled
$\ell=1\,\dots,75$ (comprising $N_{\ell}$ stocks each) according to
the first two digits of their Standard Industrial Classification (SIC)
code, which classifies major industry groups. We define a projection
matrix {\bf \sf P}, with elements $P_{\ell i} =1/N_{\ell}$ if stock
$i$ belongs to group $\ell$ and $P_{\ell i} =0$ otherwise. For each
deviating eigenvector {\bf \sf u$^k$}, we compute the contribution
$X^k_{\ell} \equiv \sum_{i=1}^N P^k_{\ell i} \,[u^k_{i}]^2$ of each
industry group $\ell$. The above procedure of computing $X^k_{\ell}$
is analogous to the analysis of wave functions in disordered systems,
where one calculates the probability of finding a particle in a given
region.

\begin{figure}[t]
\narrowtext
\centerline{
\epsfysize=0.45\columnwidth{\rotate[r]{\epsfbox{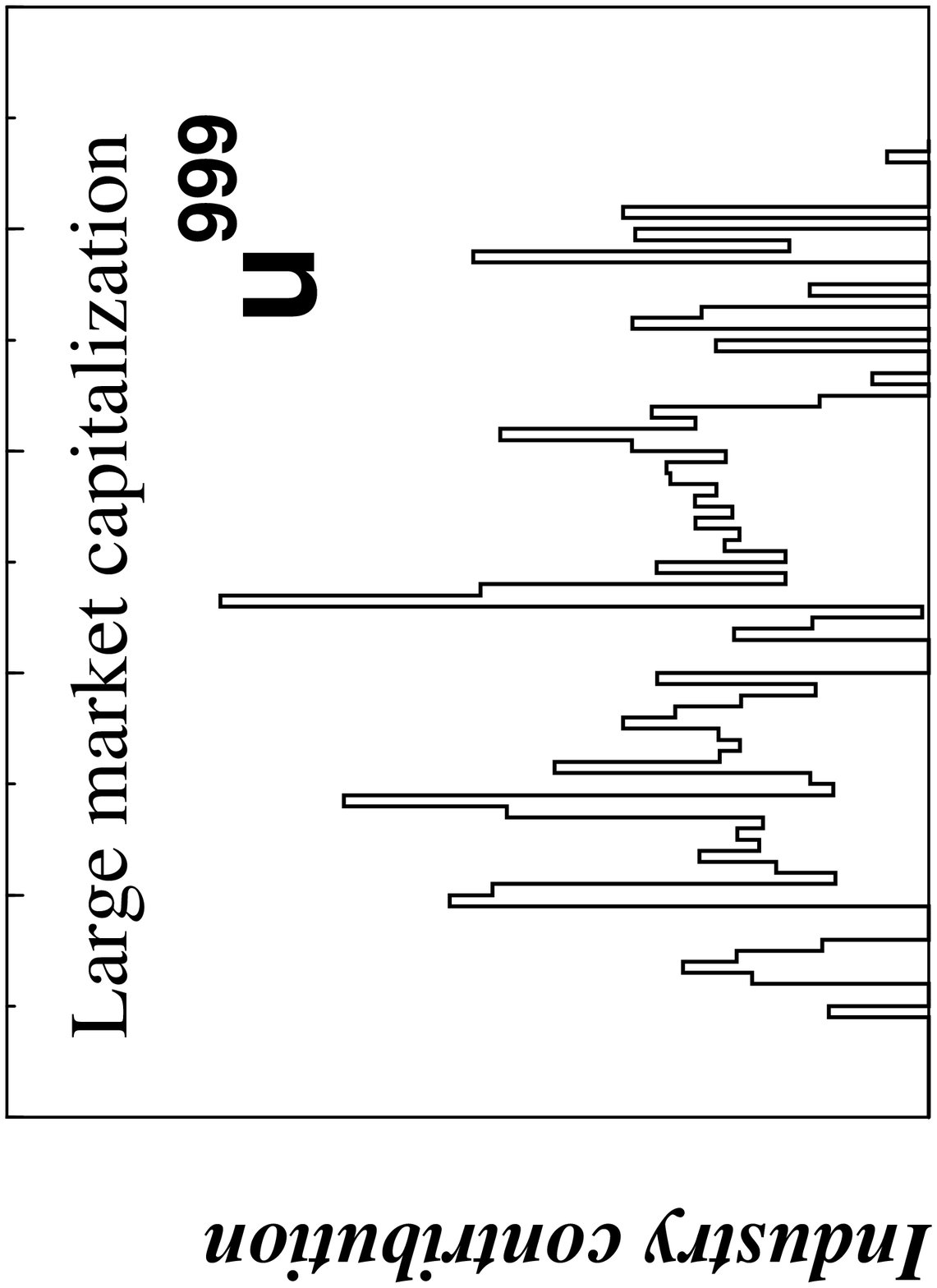}}}
\epsfysize=0.4\columnwidth{\rotate[r]{\epsfbox{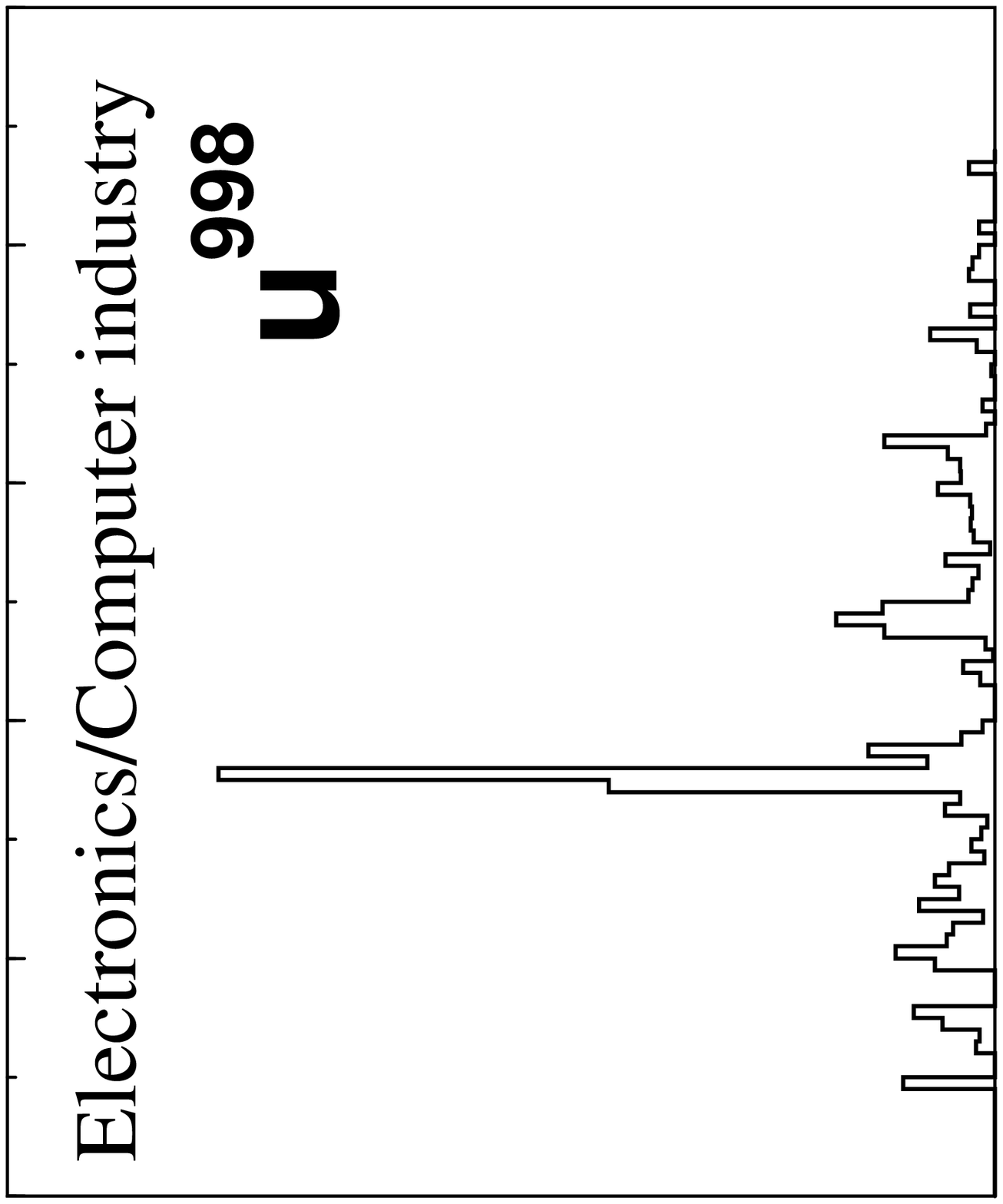}}}
}
\vspace{0.1cm}
\centerline{
\epsfysize=0.45\columnwidth{\rotate[r]{\epsfbox{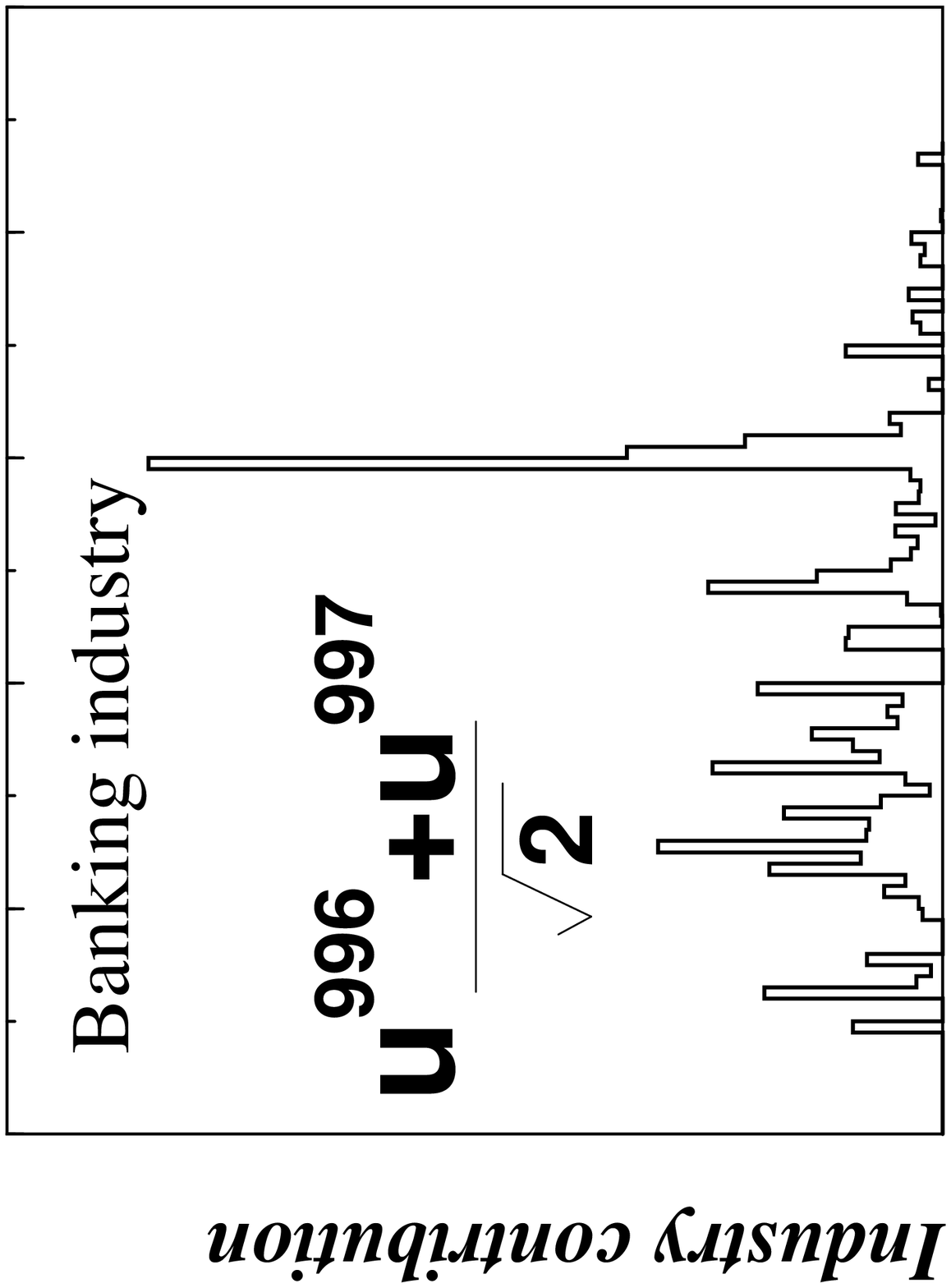}}}
\epsfysize=0.4\columnwidth{\rotate[r]{\epsfbox{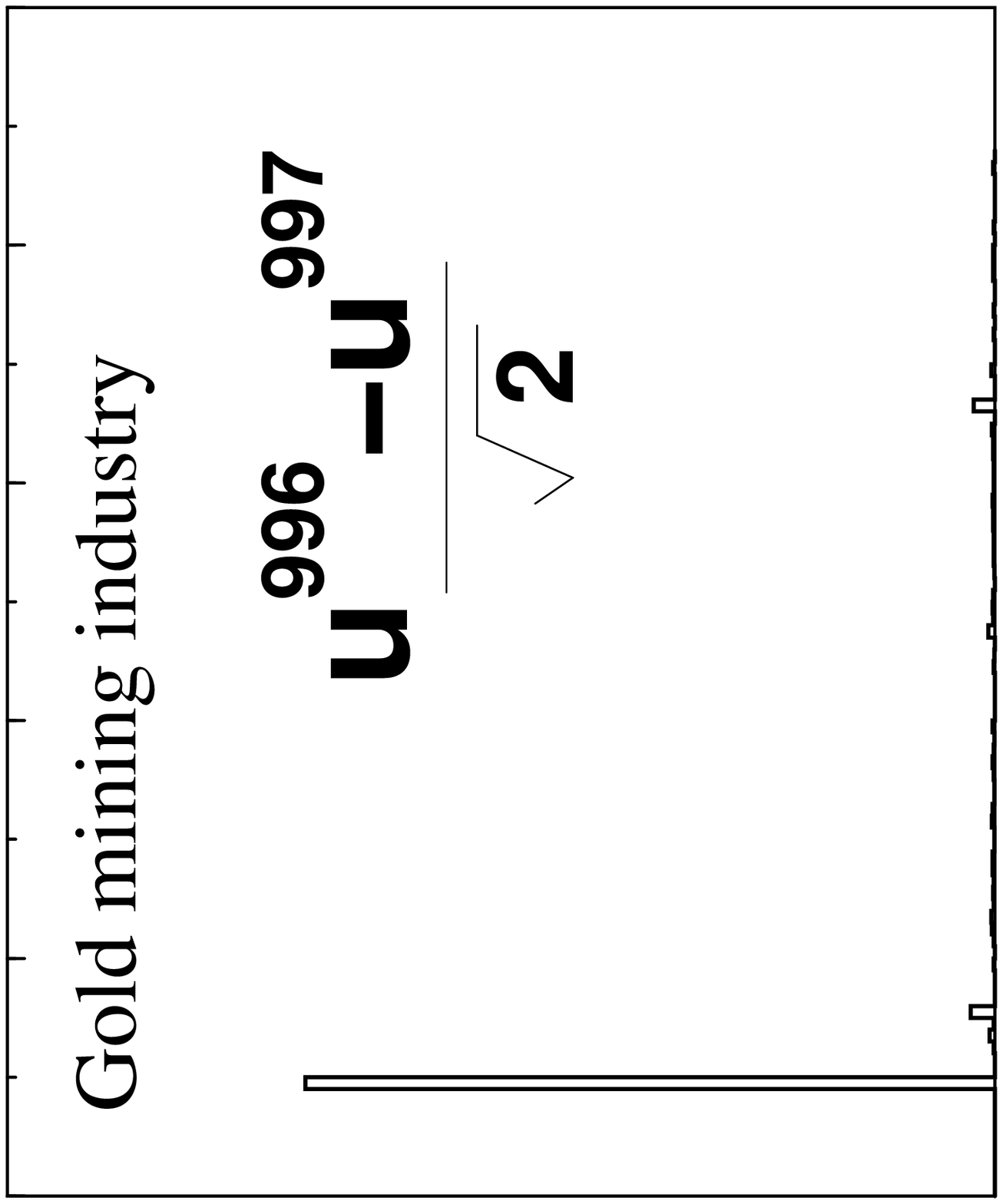}}}
}
\vspace{0.1cm}
\centerline{
\epsfysize=0.45\columnwidth{\rotate[r]{\epsfbox{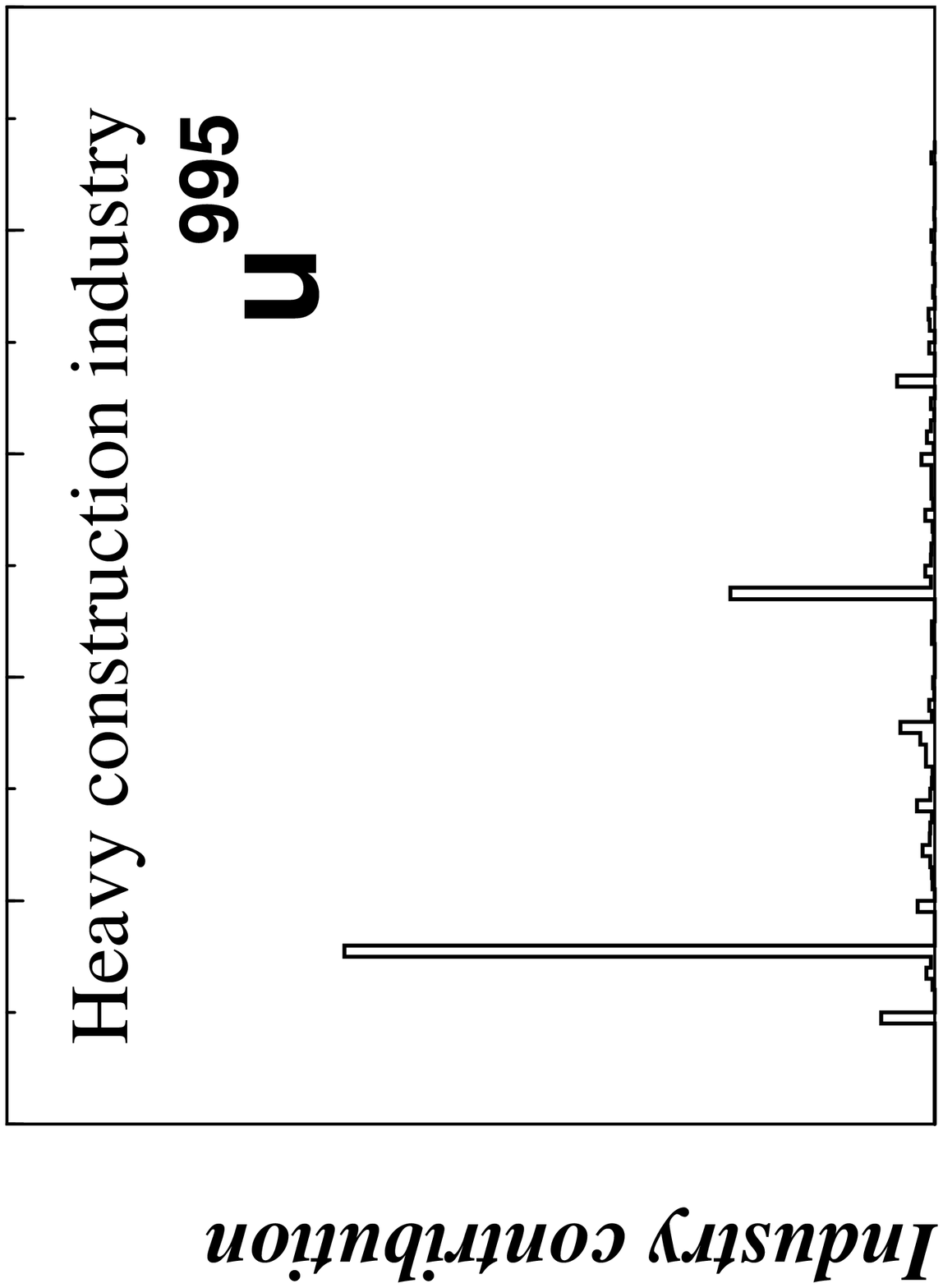}}}
\epsfysize=0.4\columnwidth{\rotate[r]{\epsfbox{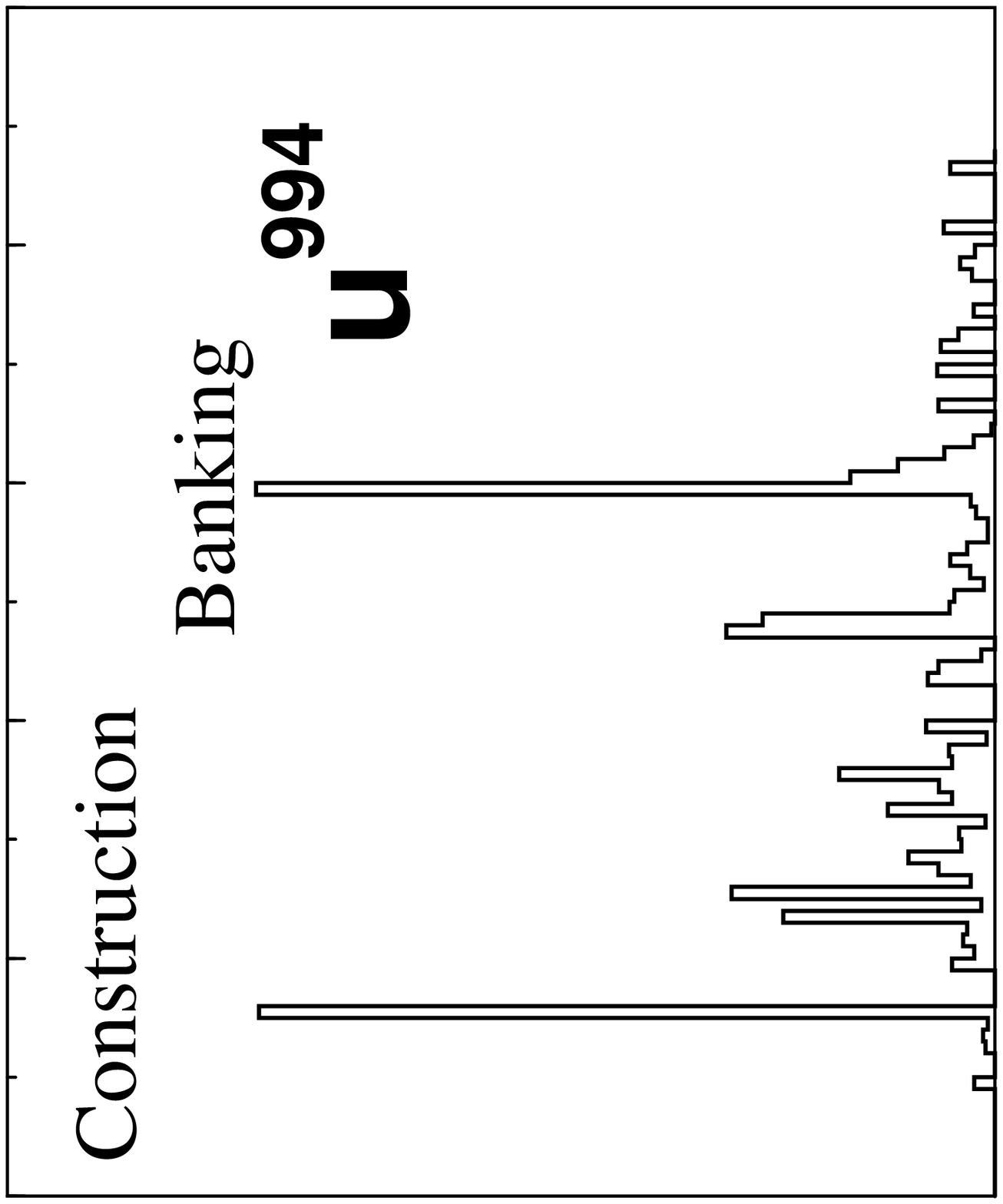}}}
}
\vspace{0.1cm}
\centerline{
\epsfysize=0.45\columnwidth{\rotate[r]{\epsfbox{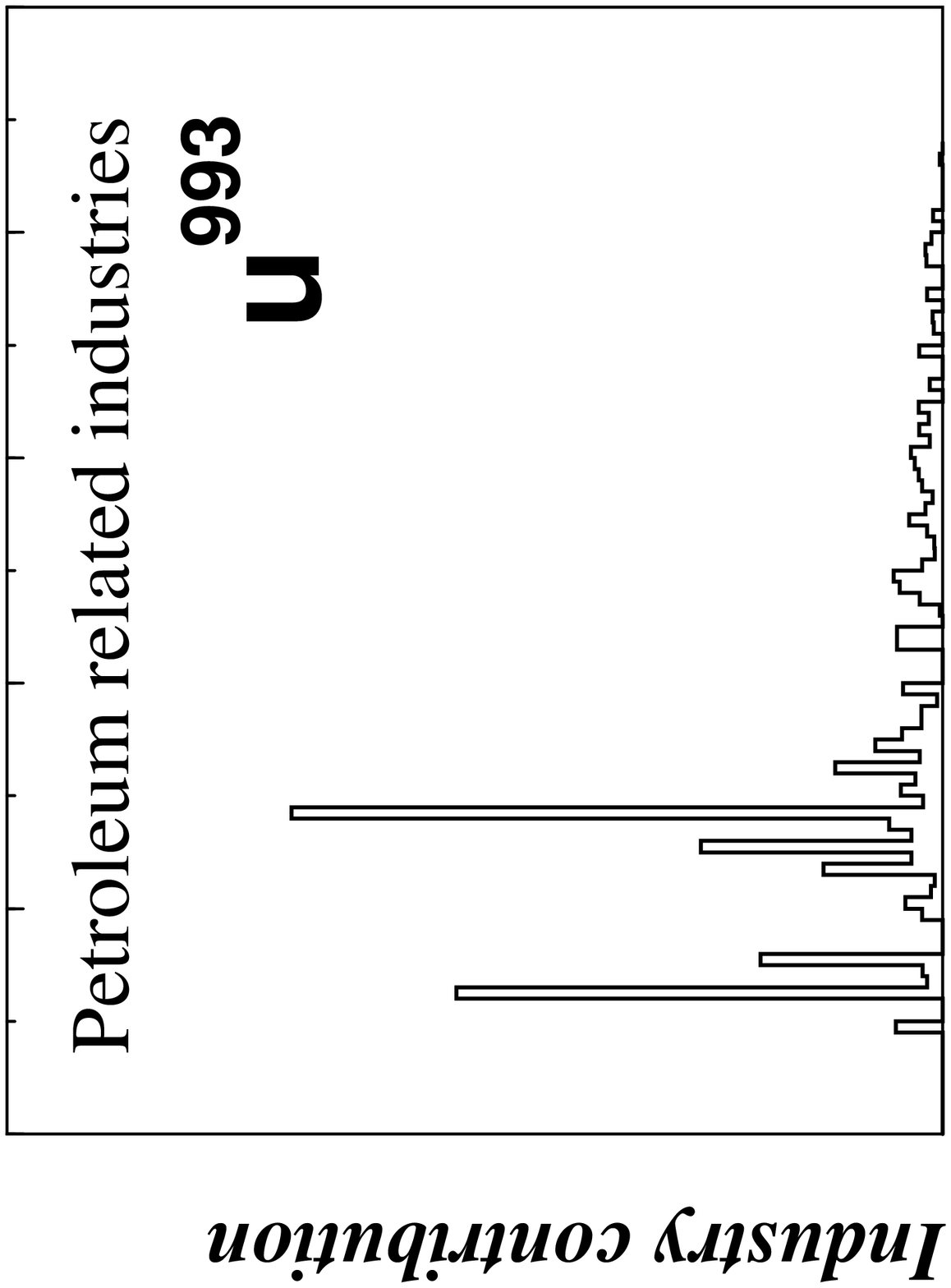}}}
\epsfysize=0.4\columnwidth{\rotate[r]{\epsfbox{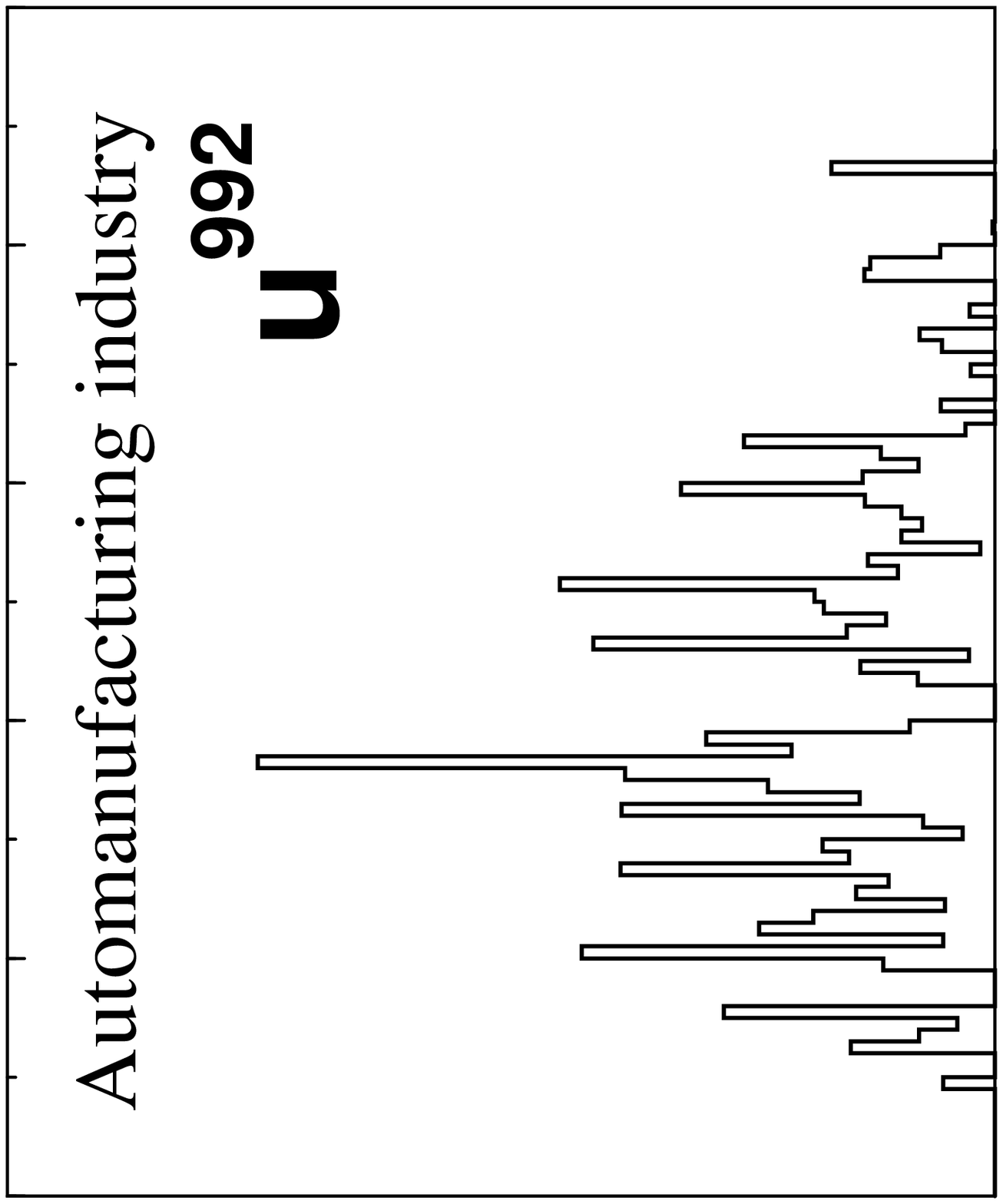}}}
}
\vspace{0.1cm}
\centerline{
\epsfysize=0.45\columnwidth{\rotate[r]{\epsfbox{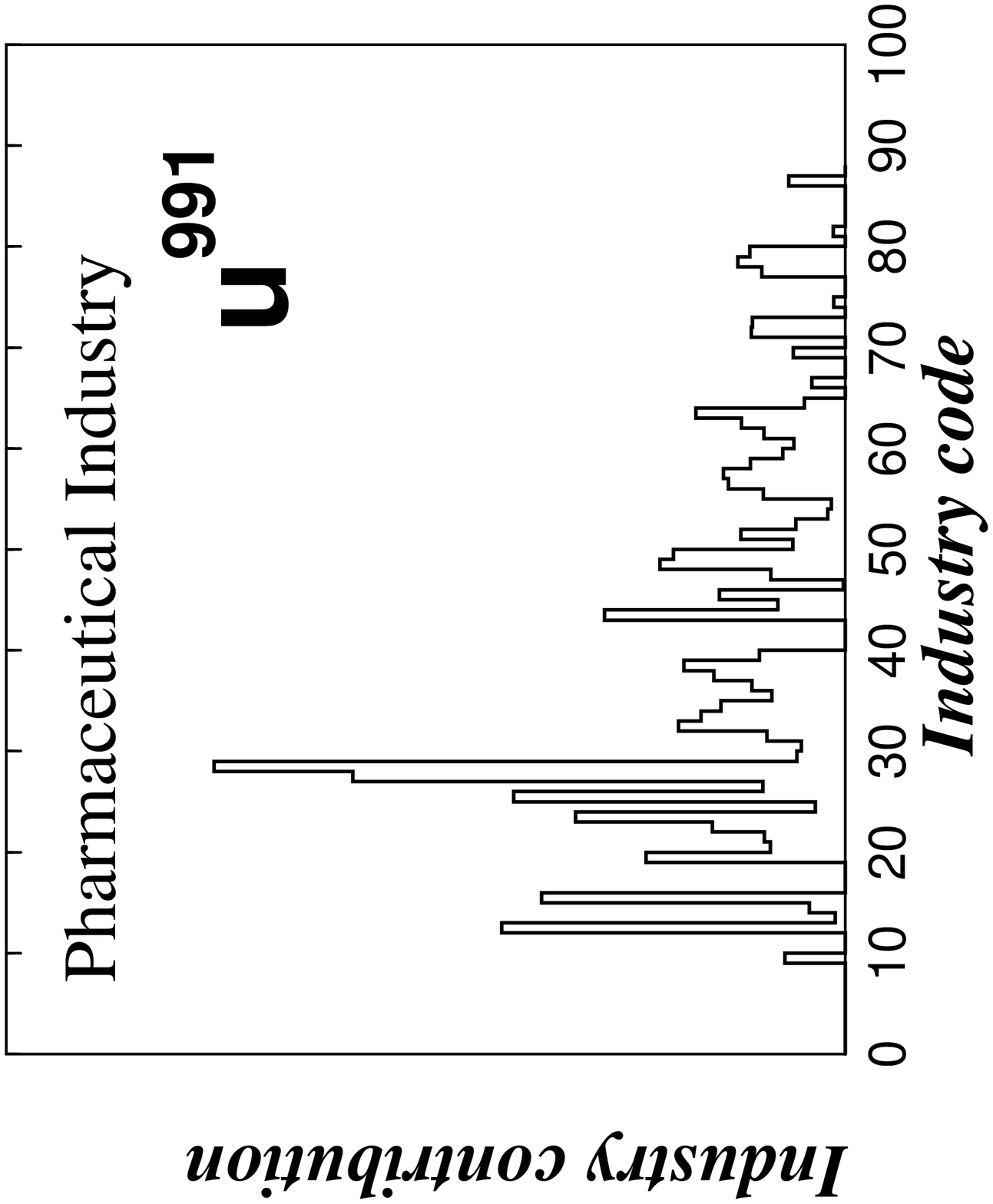}}}
\epsfysize=0.41\columnwidth{\rotate[r]{\epsfbox{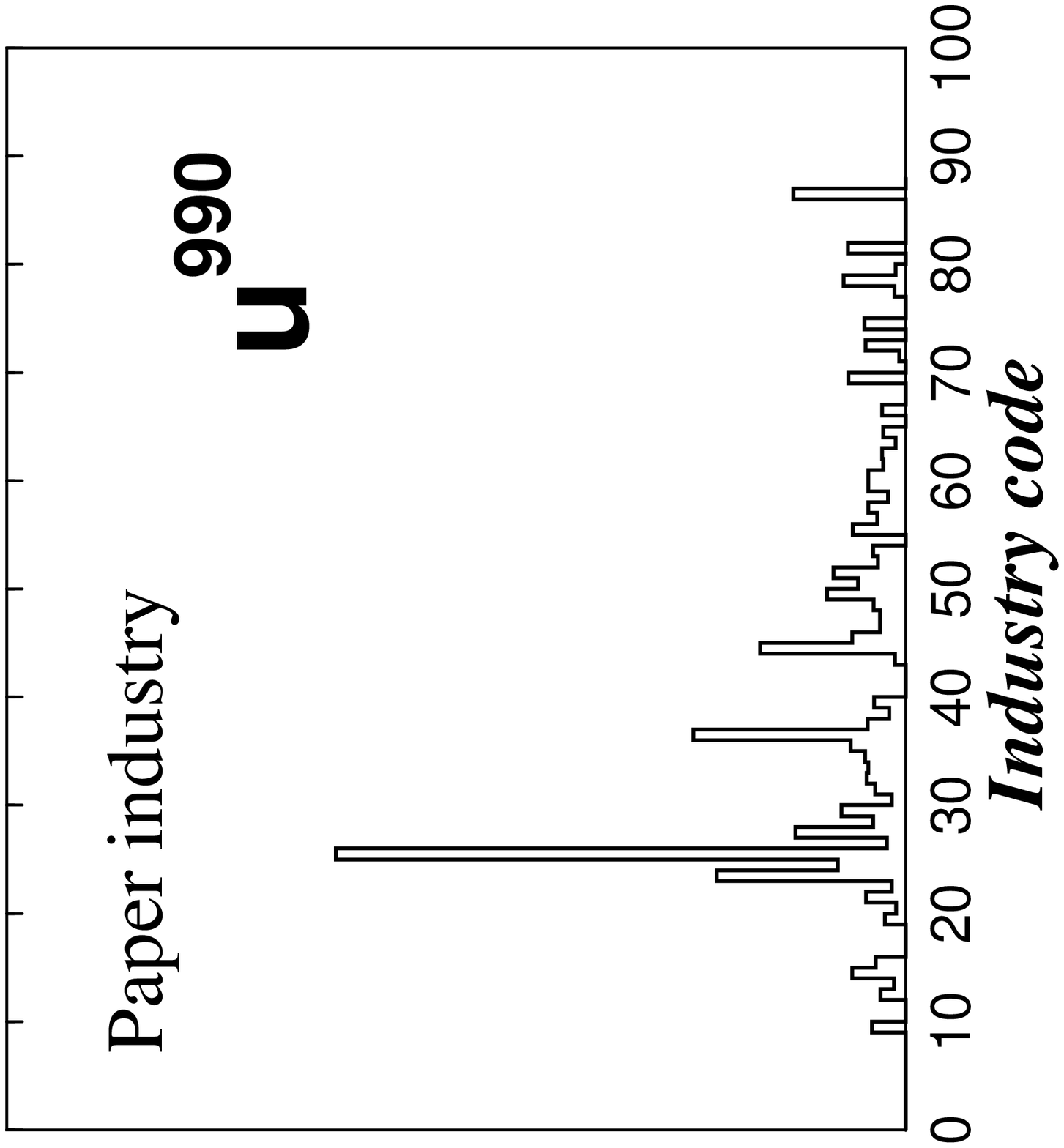}}}
}
\caption{ Contribution $X_{\ell}^k$ to industry sector $\ell$ of 
eigenvector {\bf \sf u$^k$} for the deviating eigenvectors shows
marked peaks at distinct values of SIC code, for all but {\bf \sf
u$^{999}$} which contains stocks with large capitalizations as
significant contributors.}
\label{f.projection}
\end{figure}

\begin{figure}[h]
\narrowtext 
\centerline{
\epsfysize=0.7\columnwidth{\rotate[r]{\epsfbox{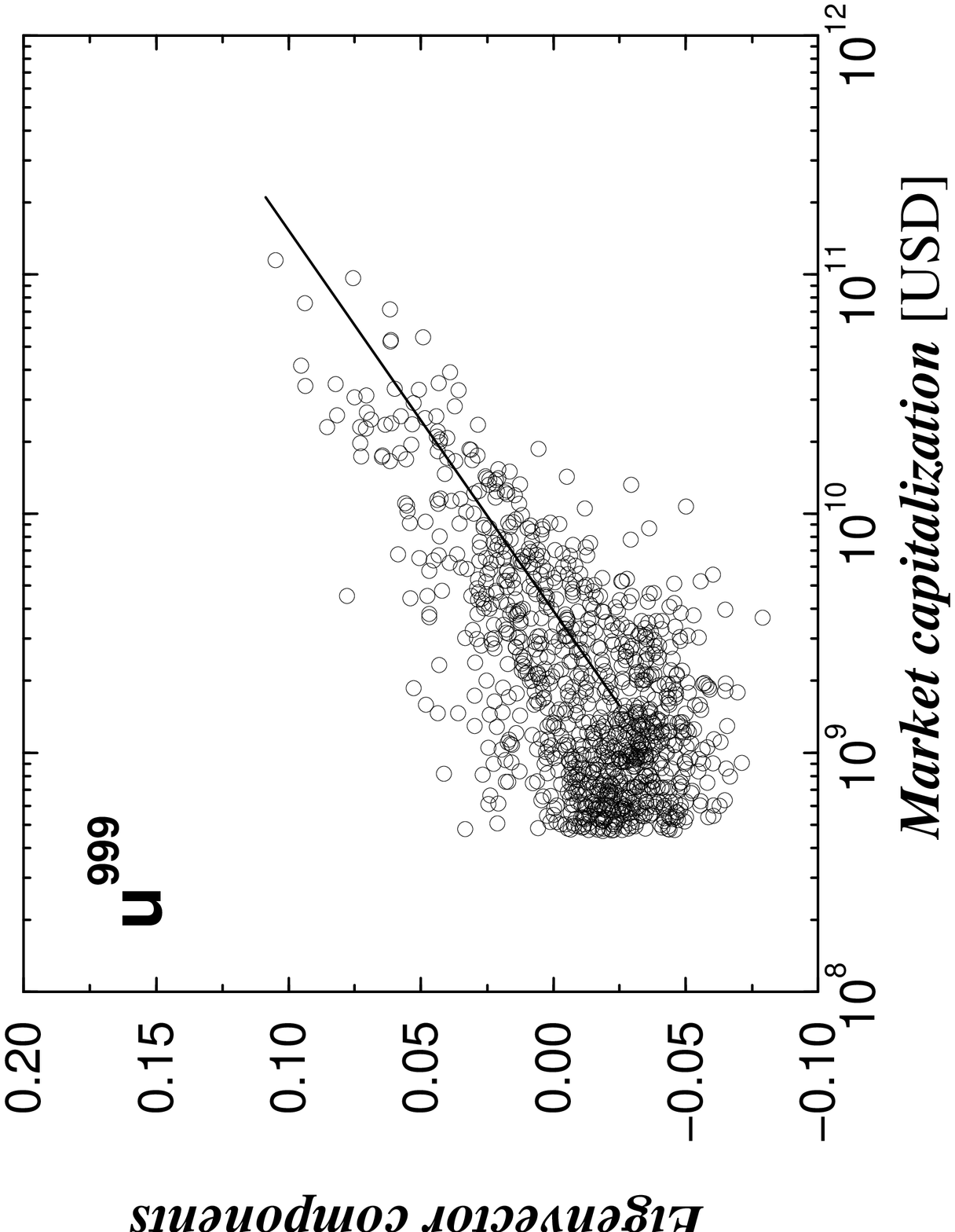}}} 
}
\caption{All $10^3$ eigenvector components of {\bf \sf u$^{999}$}
 plotted against market capitalization (in units of US Dollars) shows
 that large firms contribute more than small firms. The straight line,
 which shows a logarithmic fit, is a guide to the eye.}
\label{figmc}
\end{figure}

Figure~\ref{f.projection} shows $X^k_{\ell}$ for ten largest
eigenvectors after excluding the influence of the largest eigenvalue.
The contribution $X^{999}_{\ell}$ shows several industries.  We
examine the significant contributors and find mainly stocks with large
market capitalization [Fig.~\ref{figmc}]. We analyze $X^k_{\ell}$ for
the remainder of the deviating eigenvectors and find a significant
`peak' at distinct values of the SIC code --- suggesting that these
eigenvectors correspond to distinct industry groups~\cite{Mantegna}.

One deviating eigenvector {\bf \sf u$^{995}$} displays large values of
$X^k_{\ell}$ for firms belonging to the heavy construction industry
and telecommunications industry. In addition, an examination of these
firms shows significant business activity in Latin America. Another
case corresponds to eigenvectors {\bf \sf u$^{996}$} and {\bf \sf
u$^{997}$}, both of which contain a mixture of stocks of gold-mining
firms and banking firms.  We find that these two sectors separate when
we compute the symmetric and antisymmetric combinations {\bf \sf
$1/\sqrt{2}($ u$^{996}$ $\pm$ u$^{997})$}. The remainder of the
deviating eigenvectors display technology, metal mining, banking,
petroleum refining, auto manufacturing, drug manufacturing, and paper
manufacturing firms [Fig.~\ref{f.projection}].

\begin{figure}[h]
\narrowtext 
\centerline{
\epsfysize=0.7\columnwidth{\rotate[r]{\epsfbox{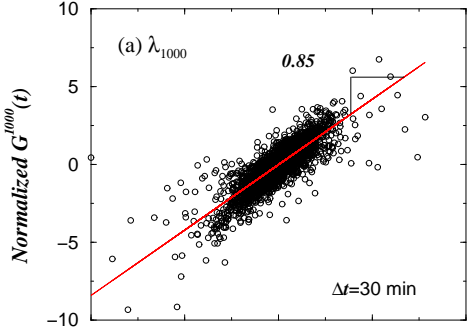}}} 
}
\caption{S\&P 500 returns $G_{SP}(t)$ regressed against the return $G^{1000}(t)$ 
of the portfolio defined by the eigenvector {\bf \sf u$^{1000}$}. Both
axes are scaled by their respective standard deviations. A linear
regression yields a slope $0.85 \pm 0.09$, showing a large degree of
correlation.}
\label{fig4}
\end{figure}

We next focus on the interpretation of the largest eigenvalue
$\lambda_{1000}$. Using the eigenvector {\bf \sf u$^{1000}$}, we
construct a time series $G^{1000}(t)\equiv \sum_{i=1}^{1000}
u^{1000}_i \, G_i(t)$. We then compare $G^{1000}(t)$ with the returns
$G_{SP}(t)$ of the S\&P 500 index, a benchmark for gauging the
performance of entire US stock market. Regressing $G^{1000}(t)$
against $G_{SP}(t)$ shows a scatter around a linear fit with slope
$0.85 \pm 0.09$ [Fig.~\ref{fig4}]. Thus, we interpret the eigenvector
{\bf \sf u$^{1000}$} as the influence of the entire market, that is
common for all stocks~\cite{Laloux99,Plerou99}.

\begin{figure}[htbp]
\vspace*{.5cm}
\hspace*{.5cm}
\epsfig{file=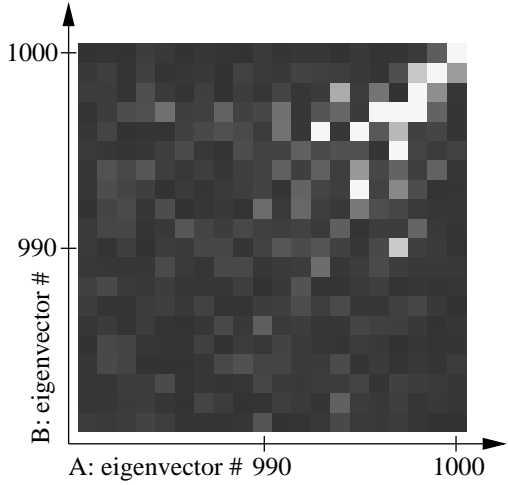,width=6.7cm,angle=0}
\caption{ Comparison of eigenvectors for different time periods A
(first half of 1994) and B (second half of 1994) by means of their
scalar product $O_{ij}$, represented on a greyscale, where zero
(black) corresponds to no overlap, and white (one) to perfect overlap.
Note that the eigenvectors corresponding to the 4 largest eigenvalues
have a large degree of time stability.}
\label{f.stability}
\end{figure}

Next, we examine whether the eigenvectors {\bf \sf u$^k$}
corresponding to business sectors remain stable in time. Partitioning
the year 1994 into two 6-month periods, A and B, we calculate the
corresponding eigenvectors {\bf \sf u$_A$} and {\bf \sf u$_B$} of the
cross-correlation matrices and quantify the time stability by
calculating the magnitude of the scalar products $O_{ij} \equiv \vert$
{\bf \sf u$_{A}^i$ u$_B^j \vert$} for the 20 largest
eigenvalues. Perfect time stability would mean
$O_{ij}=\delta_{ij}$. For $i=1000$, we find $O_{ii}=0.93$ ---
indicating almost perfect stability. We find that $O_{ii}$ decreases
as $i$ decreases from $1000$ [Fig.~\ref{f.stability}]. Extending this
analysis to daily returns using database (ii) shows that the
eigenvectors corresponding to the largest 3 eigenvalues are stable for
as many as $10$ years.

How can we explain correlations that are stable in time?  In physical
systems, one starts from the interactions between the constituents,
and then relates interactions to correlated ``modes'' of the
system. In economic systems, we ask if ``interactions'' give rise to
the correlated behavior. Interactions can arise when two companies are
doing business together, or compete for the same market. To study if
the correlations can be explained through interactions~\cite{sde}, we
model stock price dynamics by a differential
equation~\cite{Farmer,misc}, which describes the `instantaneous''
returns $g_i(t)= {d\over dt}\ln S_i(t)$ as a random walk with
interactions $J_{ij}$
%
\begin{eqnarray}
\tau_i \partial_t g_i(t)= - r  g_i(t)   + \sum_j J_{ij} g_j(t) + 
{1\over \tau_i} \xi_i(t) \ \ \,.
\label{Langevin}
\end{eqnarray}
%
Here, $\xi_i(t)$ are Gaussian random variables with correlation
function $\langle \xi_i(t) \xi_j(t^\prime)\rangle = \delta_{ij} \tau_i
\delta(t-t^\prime)$, and $\tau_i$ are relaxation times of the 
$\langle g_i(t) g_i(t+T) \rangle$ correlation function. The return
$G_i$ at a finite time interval $\Delta t$ is given by the integral of
$g_i$ over $\Delta t$.

Calculating time-dependent correlation functions for the $g_i$, we
find that correlations caused by interactions are accompanied by a
phenomenon analogous to ``critical slowing down''. The market time
series G$^{1000}(t)$ --- as well as time series constructed similarly
for other deviating eigenvectors --- have considerably larger
correlation times than a time series constructed out of a random
eigenvector, consistent with the hypothesis that correlations between
firms are caused by interactions.

The eigenvectors that we interpret as defining business sectors also
have relevance to the practical goal of finding an investment which
earns a given return without exposure to unnecessary risk (``optimal
portfolio''). Risk can be reduced by diversification of investment
into independently fluctuating groups of stocks, such as the mutually
uncorrelated business sectors that we find. Since the sectors
(eigenvectors) are stable in time, we expect the ratio of risk to
return of the portfolios constructed from them to be stable.

\begin{figure}[hbt]
\narrowtext 
\centerline{
\epsfysize=0.7\columnwidth{\rotate[r]{\epsfbox{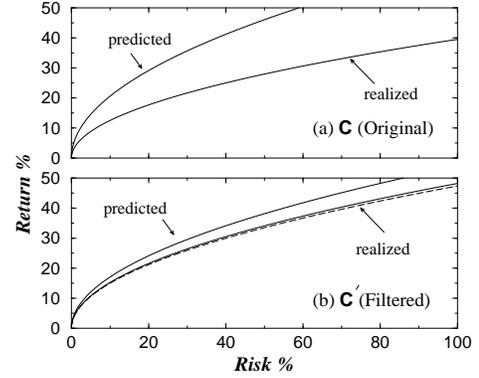}}}
}
\vspace{0.5cm}
\caption{(a) Portfolio return $R$ as a function of risk $D^2$ 
for the family of optimal portfolios (without a risk-free asset)
constructed from the original matrix {\bf \sf C}~\protect\cite{notep}.
The top curve shows the predicted risk $D_{\rm p}^2$ in 1995 of the
family of optimal portfolios for a given return, calculated using
30-min returns for 1995 and the correlation matrix {\bf
\sf C$_{94}$} for 1994.  For the same family of portfolios, the 
bottom curve shows the realized risk $D_{\rm r}^2$ calculated using
the correlation matrix {\bf \sf C$_{95}$} for 1995. These two curves
differ by a factor of $D_{\rm r}^2/D_{\rm p}^2\approx 2.7$.  (b)
Risk-return relationship for the optimal portfolios constructed using
the filtered correlation matrix {\bf \sf C$^{\prime}$}.  The top curve
shows the predicted risk $D_{\rm p}^2$ in 1995 for the family of
optimal portfolios for a given return, calculated using the filtered
correlation matrix {\bf \sf C$^{\prime}_{94}$}. The bottom curve shows
the realized risk $D_{\rm r}^2$ for the same family of portfolios
computed using {\bf \sf C$^{\prime}_{95}$}. The predicted risk is now
closer to the realized risk: $D_{\rm r}^2/D_{\rm p}^2\approx
1.25$. For the same family of optimal portfolios, the dashed curve
shows the realized risk computed using the original correlation matrix
{\bf \sf C$_{95}$}: $D_{\rm r}^2/D_{\rm p}^2\approx 1.3$.}
\label{rr}
\end{figure}

Consider a portfolio $P(t)\equiv\sum_{i=1}^{N} w_i S_i(t)$, where
$w_i$ is the fraction of wealth invested in stock $i$. The portfolio
return is given by $R=\sum_{i=1}^{N} w_i G_i$. The risk in holding the
portfolio $P(t)$ can be quantified by the variance
$D^2=\sum_{i=1}^{N}\sum_{j=1}^N w_i w_j \,C_{ij}\, \sigma_i \sigma_j$,
where $\sigma_i$ is the standard deviation of
$G_i$~\cite{Elton95}. In order to find an optimal
portfolio, we minimize $D^2$ under the constraints that the portfolio
return is some fixed value $R$ and $\sum_{i=1}^N w_i =1$. We thereby
obtain a family of optimal portfolios, which we represent by plotting
$R$ as a function of risk $D^2$ [Fig.~\ref{rr}].

To find the effect of randomness of the $C_{ij}$ on optimal portfolio
selection, we partition the time period 1994-95 into two 1-year
periods. Using the cross-correlation matrix {\bf
\sf C$_{94}$} for 1994, and $G_i$ for 1995~\cite{ack-bouch}, 
we construct a family of optimal portfolios and plot $R$ as a function
of the predicted risk $D_{\rm p}^2$ for 1995 [Fig.~\ref{rr}(a)]. For
this family of portfolios, we also compute the risk $D_{\rm r}^2$
realized during 1995 using {\bf \sf C$_{95}$} [Fig.~\ref{rr}(a)]. We
find that the predicted risk is significantly smaller than the
realized risk: $[D_{\rm r}^2 - D_{\rm p}^2]/D_{\rm p}^2\approx 170\%$.

Since the meaningful information in {\bf \sf C} is contained in the
deviating eigenvectors that define business sectors, we construct a
`filtered' correlation matrix {\bf \sf C$^\prime$}, by retaining only
the deviating eigenvectors~\cite{filter}. We repeat the above
calculations for finding the optimal portfolio using {\bf \sf
C$^\prime$} instead of {\bf \sf C}. Figure~\ref{rr}(b) shows that the
realized risk is now much closer to the predicted risk: $[D_{\rm r}^2
- D_{\rm p}^2]/D_{\rm p}^2\approx 25\%$. Thus, the optimal portfolios
constructed using {\bf \sf C$^\prime$} are significantly more stable
in time.

In summary, given only the change in price of a stock, and no
additional information about the stock, we can partition the set of
all $10^3$ stocks studied into subsets whose identities correspond
well to conventionally-identified sectors of economic activity. The
sector correlations are stable in time and can be used for the
construction of optimal portfolios with a stable ratio of risk to
return.

We thank J.-P.~Bouchaud, P.~Cizeau, E.~Derman, X.~Gabaix, J.~Hill,
M.~Janjusevic, R.~N.~Mantegna, M.~Potters, L.~Viceira, J.~Zou, and
especially L. A. N.~Amaral for stimulating discussions, and DFG grant
RO1-1/2447 for financial support.  Our results on the applications to
portfolio selection were presented at the APS March 2000 meeting by
one of us (BR) and independently by P.~Cizeau.

\end{multicols}


\begin{thebibliography}{99}
\bibitem{TAQ} 
The 30-min data analyzed are from the Trades and Quotes database
published by the New York Stock Exchange. The daily data is from the
Center for Research in Securities Prices (CRSP) database. In both
cases, only those companies that survive the entire period are
considered in our analysis.
\bibitem{sde}
Equation~(\protect\ref{Langevin}) is similar to the linearized
description of interacting soft spins. Unlike the spin model, different
stocks have relaxation times which, in general, increase by two orders
of magnitude with market capitalization --- from 10~s up to 1000~s.
\end{thebibliography}
\end{document}